# Critical success factors to improve the game development process from a developer's perspective

Saiqa Aleem [1], Luiz Fernando Capretz [1], and Faheem Ahmed[2]

[1]Department of Electrical and Computer Engineering,
University of Western Ontario, London, ON, Canada.

[2] Department of Computing Science, Thompson River University, Kamloops, BC, Canada.

E-mail: {saleem4, lcapretz}@uwo.ca, fahmed@tru.ca

**Abstract** The growth of the software game development industry is enormous and is gaining importance day by day. This growth imposes severe pressure and a number of issues and challenges on the game development community. Game development is a complex process, and one important game development choice is to consider the developer perspective to produce good-quality software games by improving the game development process. The objective of this study is to provide a better understanding of the developer's dimension as a factor in software game success. It focusses mainly on an empirical investigation of the effect of key developer factors on the software game development process and eventually on the quality of the resulting game. A quantitative survey was developed and conducted to identify key developer factors for an enhanced game development process. For this study, the developed survey was used to test the research model and hypotheses. The results provide evidence that game development organizations must deal with multiple key factors to remain competitive and to handle high pressure in the software game industry. The main contribution of this paper is to investigate empirically the influence of key developer factors on the game development process.

## 1. Introduction

The first software game was created half a century ago. In the world of software gaming, many things have changed during this time period. Now the software game industry has reached the point that it rivals other well-established industries such as music and cinema. As a result, the software gaming business has grown enormously, has made billions of dollars in profit, and has started to mature over time [1]. The game development process has also had an impact on the industry, which now counts on special methodologies and mature processes for its development, ultimately leading to an enhanced game development process. Game developers try to produce games that are different from any other game in the market. This difference can be achieved through by introducing new perspectives, new gameplays, new genre combinations, enhanced graphics, or new characters. Therefore, almost all games must be novel, and their success depends on their overall quality [2]. Only good-quality games are able to retain their players, and this has become an important factor for any software game to succeed commercially. In other words, if a game is not of good quality, players can easily switch to another game. Hence, it has become mandatory for the software game industry to try to morph and adapt to the preferences and demand of its players.

One of the main concerns in game development process is that developers need to follow best practices and procedures from software engineering discipline to develop good quality games. The game development process involves four main phases: concept, pre-production, production, and post-production [3]. It consists of various activities such as synopsis, background research, script writing, visualization and concept art, level and interaction design, animation, programming, media editing, integration, testing, and publishing. Software games are also characterized based on the category into which they fall, which is called the *genre* of the game. Genres include action, shooters, fighting, racing, adventure, sports, role playing, strategy, simulations, puzzles, dance, music, and others. Each genre has its own requirements which must be taken into consideration during the pre-production phase. For this reason, software game development is considered as a complex process that involves multidisciplinary collaborative team efforts and processes (including sound, gameplay, art, artificial intelligence, control systems and human factors) to develop a creative product. Fundamentally, game development is a form of software development process with several additional requirements such as creative design, artistic aspects, and visual presentation [4], [5]. In this context, game development organizations can apply the same software engineering principles to improve their development processes. However, many studies have discussed the challenges of applying software engineering principles to the game development process [5].

Kultima [6] highlighted these challenges from the game design perspective. Blow [4] discussed their implications from the perspective of technical frameworks and development techniques. Blow [4] and McGill [7] discussed issues even for the required technical skills for game development. Software game development also requires a range of skills that include design, project management, development, and asset creation. It also involves team members from heterogeneous disciplines, e.g., game designers, artists, programmers, and software developers. Knowledge of best practices for game development is very important and has become crucial to sustain the growth of the software game industry. Finally, this knowledge will help game developers make correct game development decisions at the right time. An investigation of key success factors from a developer's perspective will contribute to the understanding of current development process implications and will help developers improve the game development process.

Exploring diverse developers' preferences for software game development will provide a signifi-cant benefit to improve the development process by generating valuable insights. No research has been done to date on including developer-centred factors in the software game development process. This study will help identify key factors empirically from the developer's perspective, an effort that will ultimately help improve the software game development process to produce good-quality software games. To identify key factors, a quantitative survey was conducted, and the results are reported here. The survey was used to test the research model and several hypotheses. Finally, the results show that consideration of key factors from a developer's perspective helps identify important game development choices and their implications for the current process.

### 1.1 Research Background

The software game domain covers a great variety of player modes and genres [8], [9], [10]. The complexity of digital games has posed many challenges and issues in software development because it involves diverse activities in creative arts disciplines (storyboarding, design, refinement of animations, artificial intelligence, video production, scenarios, sound effects, marketing, and finally sales) besides technological and functional requirements [11]. This inherent diversity leads to a greatly fragmented domain from the perspectives of both underlying theory and design methodology. The software game literature published in recent years has focussed mainly on technical issues.

Issues of game production, development, and testing reflect only the general state of the art in software engineering. Pressman [12] stated that a game is a kind of software which entertains its users, but game development faces many challenges and issues if only a traditional software development process is followed [5],[13].

Many researchers have discussed game development challenges. Pertillo *et al.* [13] surveyed the problems faced by game development organizations. The overall game development process combines both an engineering process and the creation of artistic assets. Ramadan and Widyani [14] compared various game development strategies from a management perspective, and some researchers [15], [16], [17] have proposed frameworks for game development. To effectively manage and improve the game development, key developer's factors are required. Tschang [18] and Petrillo *et al.* [13] highlighted the issues in the game development process and its differences from traditional software development practices.

In traditional software engineering, the development phase usually involves activities like application design and implementation, and the production phase is when the software actually runs and is ready for use. However, in the game development, the production phase includes the devel-opment process, which is the pre-production phase of the software engineering process, and the production phase of software engineering is actually the post-production phase of the game development life cycle [20]. Therefore, the game development is different from the traditional software engineering process, and many researchers [5] have studied the challenges faced by this domain. Kanode and Haddad [5] stated that an important incorrect assumption has been made that game development follows the waterfall method. More recently, researchers have agreed that it must follow the incremental model because it combines the waterfall method with an iterative process. Petrillo et al. [13] reported a major concern, that developers for software creation in the game industry commonly use very poor development methodologies. The game development life cycle (GDLC) is the object of questions on many forms, which attempt to determine what types of practices are used. However, this question has no single answer. The most prominent observation made in these studies is that to address the challenges faced by the GDLC, more rigorous software engineering strategies must be used. However, the proposed GDLCs [14], [19], [20], [21] do not ensure the quality of the development process. Hagan et al. [22] published a systematic literature review of software process models used for game development. They concluded that agile and hybrid approaches are used by most organizations for game development. They also reported that Scrum [5], Kanban [23], Rapid Development Application (RAD) [24], XP [25], and incremental [5] methodologies are used by game development organizations. The major difference in software development and game development is in the design phase because design of game may undergo major change in late development. The other differences are content development and quality criteria. Managing game development has become a much harder process that anyone could have initially imagined, and because of the fragmented nature of the domain, no clear picture of its advancement can be found in the literature.

From the above discussion, it can be easily concluded that game development process is different from traditional software development process. Kasurinen et al. [26] argued that current software engineering knowledge is unable to bridge the gap between software engineering and certain aspect of game development. The overall development process to produce a game includes art, audio and gameplay other than software development discussed above. In the game development process, the content and production activities are performed in tandem with the development and engineering activities. Further, it is well agreed that the game development process is a multidisciplinary activity that involves the merging of creative and technical talent to bring a concept to life, where the main activities can be categorized into content and production, and engineering at each phase of the development process.

Moreover, sometime game development organizations reduce their development process due to of high competition and extreme market demand so they can be first to market [27]. This reduction of the development process definitely affects game quality. Therefore, they do not strictly follow the software engineering standards and practices. Because of these types of complex project-management tasks, the game development process diverges from traditional software development. Nevertheless, the differences between software engineering and games development are not exclusive; it seems that traditional software development does not fully support game development activities and provide process assessment procedures [28]. So, we need key success factors to improve game development process that may overlap with traditional software development factors or just exclusive to game development. Therefore, it has become important now to investigate the critical success factors for game development organizations in developing good quality games from developer's perspective.

### 1.2 Research Motivation

Game development has become incredibly challenging due to rapid changes in game technology such as game platforms, game engines, and reuse of code modules for different genres. During the 1990s, game development was usually carried out by small team members and involved simple architectures consisting of 2D graphics, sound, simulation, and input/output streaming. The first software games were developed by a few talented individuals from diverse backgrounds like mathematics, computer science, and physics with no educational background in engineering or computer science. At that time, developers were mainly focussed on how to develop interesting games rather than on architecture or software engineering principles. The current success of the game industry, continuous enhancements in game technology, and the need to meet the ever-higher expectations of the players resulted in a complex game development process.

The main research motivations behind this study are the rapid and continual changes in technology and the severity of competition in game development organizations. Ultimately, these factors will not only affect the business, but also will have a major impact on the game development process. Nowadays, games are developed by large teams because game projects have grown in size and complexity [4]. Various stakeholders are involved in the development process and have different expectations and world views. For example, the game designer does not know the level of complexity involved in implementing artificial intelligence to represent the behaviour of a non-player character. A software engineer may think that some features in the game design document are infeasible to implement due to time deadlines or technical constraints. Another important requirement that must be part of the game is the fun, flow, and enjoyment factors. The game develop-ment processes have different phases and are influenced by many factors. Identifying the key success factors in a game development process is extremely important for sustaining the economic growth of the software game industry.

However, very little research has been reported in the academic literature about key success factors for the game development process. Many topics in software games need attention from researchers and highlighted by some studies [22], [29], [30]. Moreover, researchers and game developers have different points of view. Basically, game developers prioritize the game development process by rapid creation and implementation of content. On the other hand, scientists and researchers prioritize investigation and research into the individual rience. This indicates that there is a need for collaboration between researchers and developers that will be ultimately beneficial to game industry standards. This study also attempts to fill this communication gap between researchers and developers. Above discussed facts, motivated us to carry out empirical investigation of key success factors that can help developers to improve their development practices. It will be ultimately enable them to develop good quality games.

The rest of the paper is organized as follows: Section 2 provides a literature review of identified factors, Section 3 describes the research methodology used for this study, Section 4 presents the results of the empirical investigation, Section 5 provides a discussion, and finally, Section 6 concludes the study.

Table 1 Identified factors from a developer's perspective

| Factors | References |
| --- | --- |
| Team Configuration & Management | Claypool & Claypool [31]; Eric et al. [32]; Musil *et al.* [25]; Tran and Biddle [35]; Stacey *et al*. [36]; Barros *et al*. [37]. |
| Game Design Document Management | Kasurinen *et al*. [38]; Bosser [39]; Callele *et al*. [40]; Callele *et al*. [41]; Reyno and Cubel [42]; Almeida and da Silva [43]; Ahmed and Jaafar [44]; Bringula *et al*. [45]. |
| Game Engine Development | Robins [3]; Sherrod [46]; Cowan and Kapralos [47]; Hudlicka [48]; Yan-Hui *et al*. [49]; Rodkaew [50]; Vanhutupa [51]; Sousa & Garlan [52]; Aitenbichler *et al*. [53]; Pimenta *et al*. [54]; Neto *et al*. [55]; Peker and Can [56]. |
| Game Asset Management | Llopis [57]; Hendrikx *et al*. [58]; De Carli *et al*. [59]; Phelps [60]; Pranatio and Kosala [61]; Lasseter [62]; Xu and CuiPing [64]; Chehimi et al. [65]; Manocha *et al*. [66]; Pichlmair [67]; Migneco *et al*. [68]. |
| Quality of Game Architecture | Wang and Nordmark [69]; Amendola *et al*. [70]; Lukashev *et al*. [72]; Rhalibi *et al*. [73]; Jhingut *et al*. [74]; Kosmopoulos *et al*. [75]; Al-Azawi *et al*. [76]; Segundo *et al*. [77]. |
| Game Test Management | Redavid and Farid [78]; Helppi [79]; Charles *et al*. [80]; Wilson [81]; Marri & Sundaresaubramanian [82]; Kasurinen and Smolander [83]; Al-Azawi *et al*. [84]; Omar and Jaafar [85]; Straat and Warpefelt [86]. |
| Programming Practices | Robins [3]; Sarinho and Apolinario [87]; Czarnecki and Kim [88]; Chen *et al*. [89]; Anderson [90]; Xu and Rajlich [91]; Zhang *et al*. [92]; Wang and Norum [93]; Meng *et al*. [94]. |

components of a system. Researchers do not have resources to develop a standard game, whereas developers never publish the results of their expe-

2. **Literature review and proposed hypothesis**

In recent times, the Software Game Industry (SGI) has seen unprecedented growth. To succeed in a highly competitive environment, game developers must bring innovative, good-quality games to the table. Identifying key success factors to improve the game development process will help developers maintain the pace. Key factors in the game development process are the least addressed area in software game research. Various factors have been identified from a literature review of published articles on software games as a basis for discussion of the game development process.

Table 1 briefly presents the identified factors, with references for each. The identified factors and the related literature are described in the following sub-sections.

### 2.1 Team Configuration and Management

The development of software games involves multi-disciplinary team configuration and management. More specifically, team configuration and management are considered critical to the success of any game development project. Game development requires intensive team management [31]. Team management can be defined as the process of administration and coordination between groups of individuals who are performing specific tasks [32]. It involves forming different groups, establishing collaboration among them, setting objectives for a common set of interpersonal dynamics among team members, and performance appraisals. The game development process also involves configuration and management of multidisciplinary teams or teamwork projects and management of the collaboration among them. The term "teamwork" refers to group of individuals who are completing a specific task [33].

The term "collaboration" can be defined as the level of shared understanding and coordination among teams and the maintenance of this level [34]. Very few research studies have investigated the importance of multidisciplinary team configuration and management in software game development. Musil *et al*. [25] highlighted the importance of heterogeneous team collaboration in the video game development process. They proposed a method based on the Scrum methodology to improve workflow integration and collaboration between heterogeneous game development team members. The proposed process separates the pre-production, production, and post-production phases. Management through collaboration and integration of heterogeneous disciplines in game development is achieved by executing daily heterogeneous discipline-specific workflows in a sprint iteration adjusted by daily scrums. They claimed that this approach will enable each discipline to use the workflows in which they are most proficient in accordance with the demands and pace of other involved disciplines.

Tran and Biddle [35] discussed the collaboration factor for team management in serious game development. They explained that the collaborative process is based on ethnography and a qualitative approach. The proposed model includes many factors such as physical resources, social relationships, organizational goals, and team knowledge. They conducted a case study that determined that collaboration between multidisciplinary team requires teams to communicate frequently, to respect each other's contributions, and to share the same model and goals for game development. Stacey *et* *al*. [36] and Barros *et al*. [37] also investigated the collaboration factor in multidisciplinary game development teams and the development of computer games.

To determine whether proper team configuration and its management has any impact on the game development process, "team configuration and management" was selected as an independent variable, as shown in Fig. 1. Hence, Hypothesis 1 and corresponding null hypothesis can be stated as follows:

**Hypothesis 1:** Team configuration and management have a positive influence on the enhanced game development process.

**Null Hypothesis:** Team configuration and management have no influence on the enhanced game development process.

### 2.2 Game Design Document Management

The Game Design Document (GDD) has also been identified as an important factor in improving the game development process. The GDD is the outcome of the pre-production phase of game development. It is developed and edited by the game design team to organize their efforts and their development process. The form of the GDD varies widely across studios and genres. Basically, the GDD includes the goals of the game, the genre of the game, the overall flow, the story behind the game, the characters and their dialogue, special effects, the number of elements and feature fits within the game, and feature creeping information if required. Typically, this document is developed to express the concept of the game and to provide a basis for requirements engineering in the game development process. Game designers can trace back all their efforts to the requirement analysis in the GDD.

In the game development process literature, researchers have explored the importance of the game design document and its management in various ways. Some of them have highlighted the importance of the GDD by discussing the importance of requirements engineering in game development. For example, Kasurinen *et al*. [38] highlighted the importance of requirements engineering in the game development process. They interviewed 27 software professionals from game development organizations to obtain insight into their development process. The findings of the study showed that the professionals follow approaches or methods that are somewhat comparable to requirements management and engineering, but not to particular requirements engineering practices. Bosser [39] suggested that massively multi-player game design needs a prototyping tool and proposed a framework model to facilitate its design. They also suggested that game prototyping is important and helpful for better game design. Callele *et al*. [40] also investigated the importance of requirements engineering in the video game development process. They suggested that the reasons for the failure of any game may be rooted in problems of transforming the pre-production phase document, i.e., the GDD, with any implied information and application of domain knowledge from the pre-production phase into the production phase.

An understanding of upcoming media and

technology developments, game play, and non-functional requirements is also considered important for the GDD. Callele *et al*. [41] described how the GDD is helpful in obtaining a better understanding of the game design process and explained the definition of gameplay process in cognitive game development. Reyno and Cubel [42] proposed a model-driven game development method that ultimately accelerates game design. Almeida and da Silva [43] performed a systematic review of game design methods and of various available tools. They emphasized the use of standardized tools to develop the GDD. Other researchers have emphasized inclusion of the user perspective and have provided game design guidelines. Ahmed and Jaafar [44] emphasized the importance of user-centered game design and proposed that it should be considered at the concept phase of game development. Bringula *et al*. [45] gathered user perceptions to determine how a serious game should be developed. Based on their study, they suggested some design guidelines for four-dimensional game design, including storyline, aesthetics, reward systems, and the game objective.

To develop a good-quality game, the GDD must be properly managed so that production team members can easily move it into game production. GDD management has also been selected as an independent variable in this study, and therefore the following hypothesis and corresponding null hypothesis are proposed:

**Hypothesis 2:** Proper management of the game design document has a positive and significant effect on the overall game development process.

**Null Hypothesis:** Proper management of the game design document has no effect on the overall game development process.

### 2.3 Game Engine Development

Game engines are considered to be a powerful tool by game developers and have been in use for more than two decades. A game engine is a software layer that helps in the development process by enabling developers to focus solely on game logic and experimentation [3]. Many commercial game engines are available to help game developers with advanced rendering technologies and code reuse, resulting in shorter development time and reduced cost. Sherrod [46] defined the game engine as a "framework comprised of a collection of different tools, utilities, and interfaces that hide the low-level details of the various tasks that make up a video game". Overall, the game engine represents the basic structure of the game as it appears in the middle layer, between the application layer and the various underlying platforms.

In the literature, most researchers often use the terms "game engine" and "game development framework" interchangeably. This study uses the term "game engine" to refer to the development tool that includes most of the functionality and features that become part of any software game. The list of primary features that can be part of any modern game engine includes scripting, rendering, animation, artificial intelligence, physics, audio, and networking. Cowan and Kapralos [47] performed a survey on frameworks and game engines for serious game development only. They compared all the commercially available game engines and their various features. The results of their study suggested that most of the game engines that have been developed to create entertainment games can also be used for serious game development. Hudlicka [48] suggested a set of requirements that are necessary for game engine development, specifically for affective games. Research has been also done on development of game engines specific to different platforms, such as for the Android platform [49], a 3D role-playing game for cross-platform development [50], and the Browser games [51].

A few researchers have explored the means of addressing the challenges faced by developers in supporting and building development tools [52], [53]. However, they were not successful in achieving the required feature and design flexibility. Researchers proposed different solutions for game engines to address the challenges they faced. Pimenta *et al.* [54] proposed that game engines enable fast learning for game developers and include the ubiquitous characteristics of the game design and development process. Neto *et al.* [55] discussed the issue of game engine standardization in software game development. Game developers are interested in producing the same game for different platforms and rely mostly on the same game engine. They suggested that commonality and variability assessment must be done to enable game engine reuse. Peker and Can [56] proposed a methodology for developing game engines for mobile platforms based on design goals and design patterns. They emphasized the need to design goals and strategies for implementation in the game engine. For mobile platforms, the basic design goals suggested by them were usability, efficiency, portability, and adaptability. To determine whether standard game engine development has a positive impact on the overall game development process, game engine development was considered as an independent variable in this study. Hence, the following hypothesis and corresponding null hypothesis are proposed:

**Hypothesis 3:** Game engine development has a significant impact on the game development process.

**Null Hypothesis:** Game engine development has no impact on the game development process.

### 2.4 Game Asset Management

Anything can be considered as a game asset that contributes to the visual appearance of a game, whether artwork (including 3D elements or textures), music, sound effects, dialogue, text, or anything else. Llopis [57] stated that "game assets include everything that is not code: models, texture, materials, sound, animations, cinematics, scripts, etc." Actually, game assets include any piece of data that can be used by a game engine aside from code, scripts, and documentation. The elementary unit of game assets can be referred to as a game bit [58] and typically has no value when considered independently. There are two categories of bits: characters, which can be an asset that interacts in a simulated environment, and abstract bits, which are kinds of sound and texture that can be use together to produce a concrete bit. The main six kinds of game bit are texture, sound, vegetation, buildings, fire, water, stone, clouds (concrete), and behaviour. Game space definition is another game asset, which is part of content generation for any game. It provides a kind of game environment where game bits can be placed.

In the literature covering game asset creation and management, researchers have explored game assets in term of animation, audio processing libraries for different genres, and content generation for games. De Carli *et al.* [59] and Hendrikx *et al.* [58] carried out a survey of procedural content generation techniques for game development. Animation in games is considered an important asset because it has a great impact on game performance [60]. Studies have been done to explore animation models for different genre of games. Pranatio and Kosala [61] performed a comparative study of keyframes [62] and skeletal animations [66] for multiplayer games. Their results indicated that skeletal or bone-based frames are better than keyframe models in term of memory load and frames per second. Xu and CuiPing [64] reviewed currently used 3D accelerators for graphics animation. A wide variety of graphics cards are available to programmers. Hence, they discussed the current benefits and limitations of APIs such as OpenGL and DirectX. Chehimi *et al.* [65] described the evolution of 3D graphics for mobile platforms. They concluded that the current market presents challenges regarding graphics quality and battery life of mobile devices. These need to be addressed by standardizing successful game development for mobile platforms.

Sound within a game is one of the game assets that enable developers to build responsive, interactive, and attractive games. Currently, game development relies on pre-recorded sound clips that can be triggered during any game event [66]. These can be managed through dynamic audio processing libraries. Researchers have also studied the use of audio processing libraries in software game development. Pichlmair [67] studied music games and determined that they can be classified into two categories, rhythm and instrument games. Their analysis showed that music in video games has seven qualities: rhythm, active score, quantization, synesthesia, play as performance, sound agents, and free-form play. Migneco *et al.* [68] proposed an audio processing library to enable use of sound in Web-based games using a Flash development tool. They claimed that this approach provided flexibility and great functionality for developing games using Flash technology.

For the reasons discussed above, creation and management of the number of assets required for game development has become challenging. Mechanisms are needed to control the different versions of assets that are developed for games. Commercially, a number of tools are available, such as 3D Studio Max, Maya, and Adobe Photoshop, which can also create various assets like textures, 3D models, animations, sound effects, music, voice recordings, levels, and scenes. Modern game engines also include modules for asset management. Based on a literature review of game asset management, this study has considered game asset management as another independent variable that is considered important for the game development process. Hence, the following hypothesis and corresponding null hypothesis are proposed:

**Hypothesis 4:** Game asset management is important for enhancing the game development process.

**Null Hypothesis:** Game asset management is not important for enhancing the game development process.

### 2.5 Quality of Game Architecture

The primary function of game architecture is to support game play. It helps to define challenges by using constraints, concealment, exploration, and obstacles or skill testing. Game architecture is a kind of blueprint for the underlying complex software modules. It is used to delineate design, perform trade-off analysis, and investigate system properties before implementation and potential reuse. Basically, it draws together gameplay factors and technical requirements. A perfect game architecture would have modularity, reusability, robustness, and tractability features.

The importance of software architecture in game development has rarely been researched. Only Wang and Nordmark [69] have explored this topic. Their finding was that software architecture plays an important role in game development, with the focus mainly on achieving high performance and modifiability. They also stated that most developers use game-specific engines, middleware, and tools for game development. A number of studies have explored these various development frameworks. The proposed game development frameworks can help game developers to define their game architecture. Amendola *et al.* [70] proposed a framework for experimental game development called GLIESE. They proposed that a game architecture should have at least three sub-systems: a game logic processing system (view and model), a graphic processing system (graphic interface and view interface), and an input processing system (event manager, controller, and event publisher). These sub-systems must be clearly separated so that they can work independently. The authors suggested a Model-View-Controller (MVC) [71] pattern for the architecture. Basically, this pattern divides the application into three components: model, view, and controller. The defined relations and collaboration among these components helps in game deployment because ultimately the code associated with each sub-system's logic will operate in the desired manner.

Lukashev *et al.* [72] proposed a mobile platform development framework specifically for 3D application. They claimed that their suggested approach will help developers improve the development process. The first stage of the proposed framework is the design phase for creation of the initial model (2D or 3D) and selection of the right modeling tool and graphic format. The second stage, the integration stage, enables developers to put together already-created models into scenes and create animation. The authors suggested that a structural optimization technique can be used to create scenes. The next stage is the utilization stage, in which the created models are converted to mobile format. Implementation is the final step of the framework, where developer put together source code, auto-generated source code, and created resources. Several other studies have also been performed to propose development frameworks for various platforms based on different technologies for defining the system architecture. For example, Rhalibi *et al.* [73] proposed a 3D

Java framework for Web-based games, Jhingut *et al.* [74] and Kosmopoulos *et al.* [75] proposed a framework for mobile platforms, Al-Azawi *et al.* [76] proposed an agent-based agile methodology for game development, and Segundo *et al.* [77] proposed a game development framework specific to the Ginga middleware.

From the preceding discussion, it is clear that the quality of the game architecture is important for the game development process, and therefore it was considered as another independent variable in this study. The following hypothesis and corresponding null hypothesis are therefore presented:

**Hypothesis 5:** Quality of game architecture has a positive impact on the enhanced game development process.

**Null Hypothesis:** Quality of game architecture has no impact on the enhanced game development process.

### 2.6 Game Test Management

Game testing is a very important phase of game development. A game can be tested at different levels of development because game testing is different from software testing [78]. There are many steps involved in game testing other than test-case definition because most game testing is based on black-box testing. Hence, management of overall game test methods becomes crucial. In the pre-production phase, a test plan document should be established to set standards for the game software. Game quality can be evaluated according to the graphics, sounds, and code that are compiled into the game code. Proper documentation of testing helps developers fix problems more quickly and cheaply. Delays in testing can result in project failure.

Helppi [79] discussed many game test methods that can be used during the development phase, such as smoke testing that is used to test the user interface logic. Regression testing is performed to check that game quality is still good after any change such as addition of features or add-ons. Connectivity testing is used for networking games and mobile games to test client-server interaction. Performance testing can ensure the real performance of the game. Abuse testing is performed by giving multiple inherent inputs through the controller and determining game performance. Compliance testing makes sure that any compliance standards enforced by any stakeholder are met. Finally, functional testing verifies overall game play and reveals issues related to stability, game flow, game mechanics, integration of graphic assets, and user interface. Redavid and Farid [78] also discussed game testing methods used to detect interactions failures and listed them under the term combinatorial testing [79]. The second approach involves test flow diagrams, which are used to develop models of game behaviour from a player's perspective. Third is cleanroom testing, which helps to determine game reliability. The test tree is another testing method discussed by the authors, which can used to organize test cases.

Wilson [81] also argued that no one testing method is better than another. He suggested that good testing is a combination of 30% of ad-hoc testing, 40% test cases, and 30% alternating between the two until the strengths of both are de-termined. Marri and Sundaresaubramanian [82] discussed game test methods and suggested that the game tester should test game quality by verifying game play, logical consistency, observability, progressive thinking, and reasoning ability, as well as exhaustively testing features, game strategy, and functionality. Kasurinen and Smolander [83] interviewed seven game development teams from different organizations and studied how they test their games based on grounded theory. They concluded that all participating organizations had the resources to perform technical testing, but that they relied mostly on exploratory and usability testing rather than using a pre-planned approach. Al-Azawi *et al.* [84] proposed a set of evaluation heuristics that could be used in game development methodologies for most game genres. Omar and Jaafar [85] proposed a tool to evaluate the usability of educational games, and Straat and Warpefelt [86] suggested use of the two-factor theory to evaluate game usability.

Management of game testing during the game development process has clearly come to be of crucial importance for game developers. Hence, test management was selected as another independent variable in this study, and the following hypothesis and corresponding null hypothesis were proposed:

**Hypothesis 6:** Game test management has a positive impact on the enhanced game development process.

**Null Hypothesis:** Game test management has no impact on the enhanced game development process.

## 2.7 Programming Practices

Good programming practices are a very important factor in successful game development. A programming team with the necessary skills is definitely considered as the backbone of the game development process. The programmer must select the right coding architecture for each game project. Basically, the lead programmer must select between two types of coding style: either game-specific code (the programmer has to develop everything by him/herself) or game-engine code (where the game engine is the foundation for a game-specific code). The game code can then be organized in various ways [3], such as an ad-hoc architecture where the programmer must deal with tightly coupled code. Another choice is a modular architecture-based coding style, where the programmer identifies and separates the code into different modules or libraries. In this type of programming, reuse and maintainability are improved over ad-hoc-based coding. However, dependencies between different modules cannot be controlled, which may lead to tight coupling. The directed acyclic graph (DAG) is another way of organizing code. This is also a modular architecture-based coding scheme in which dependencies between modules are tightly controlled. Layered-style coding is also based on a DAG architecture, but modules are arranged in rigid layers, and each can interact only with the modules in the layer directly below.

Game programming involves a wide range of issues and considerations. Most researchers have tried to address these individually. The first is the issue of coupling between different modules. Sarinho and Apolinario [87] tried to address this problem using a proposed generative programming approach. Generative programming aims to automate the software development process using a number of static and dynamic technologies including reflection, meta-programming, and program and model analysis [88]. The proposed method was based on a game feature model that could represent both common and variable implementation aspects of software games. Meta-programming resources were used to generate and represent compatible source code for available game frameworks and game engines. The authors concluded that the proposed approach would result in loss of the coupling development strategy between game implementation and its domain software artifacts. Code cloning in open-source games is another issue discussed by Chen *et al.* [89]. They provided a detailed study of the issues of code clones in more than twenty open-source game projects based on C, Python, and Java for various game genres. Selection of a scripting language is another issue in game programming. Anderson [90] discussed the classification of scripting systems used for software games. Xu and Rajlich [91] described a study that explored pair programming practices and concluded that paired programmers completed their task faster with higher quality. They suggested that pair programming is a good approach for game development.

Selection of a programming language is another challenge for today's game developers. Many studies have been done to explore different programming languages for different platforms.

Zhang *et al.* [92] performed experiments on five industrial RPG mobile games developed using the object-oriented programming paradigm. Optimization strategies with structural programming were applied to the same code. The results of the study showed that object-oriented programming must be used with great care and that structural programming is also a good option for mobile game development. Another study [93] highlighted the issues for game development posed by wireless peer-to peer games in a J2ME environment using an available Bluetooth API. The issues discussed included slow device discovery, Bluetooth transfer speed, extra resource consumption, and Bluetooth topology. Meng *et al.* [94] developed a peer-to-peer online multiplayer game using DirectX and C# to achieve playability in a .Net environment.

According to the above discussion, programming practices were selected as an independent variable in this study, and the following hypothesis and corresponding null hypothesis were proposed:

**Hypothesis 7:** Good programming practices are important for the enhanced game development process.

**Null Hypothesis:** Good programming practices are not considered important for the enhanced game development process.

## 3. Research Model

The main objective of the proposed research model is to analyze the interrelationship between key factors and game development and also to understand the influence of these factors on overall game quality in the SGI market. The model's theoretical foundation is based on existing concepts found in the game development literature. Note that most studies in the literature discuss one or two of the factors mentioned above for software games and their impact on the overall game development process. To the best of the authors' knowledge, this is the first study in the game development literature that highlights key factors in game development. This study proposes to investigate empirically the influence and association of key game development factors. Figure 1 presents a theoretical research model used in this study, which will be empirically investigated. The theoretical model evaluates the relationships of various independent variables emerging from software engineering and management concepts such as project management, theory, and behaviour with the dependent variable, enhanced game development, in the context of the game development process. This study mainly investigates and addresses the following research question:

**Research Question:** How can game developers improve the game development process?

The research model includes seven independent variables: team configuration and management, game design document management, game engine development, game asset management, quality of game architecture, game test management, and programming practices, and one dependent variable: the enhanced game development process.

The multiple linear regression equation of the model is given as Equation 1:

Enhanced game development process =

$\alpha_0 + \alpha_1 f_1 + \alpha_2 f_2 + \alpha_3 f_3 + \alpha_4 f_4 + \alpha_5 f_5 + \alpha_6 f_6 + \alpha_7 f_7$,

Where $\alpha_0, \alpha_1, \alpha_2, \alpha_3, \alpha_4, \alpha_5, \alpha_6, \alpha_7$ are coefficients and $f_1$–$f_7$ are the seven independent variables.

## 4. Research Methodology

Developing a software game involves phases such as pre-production, production and post-production, in which each phase contains a number of activities. Some of these activities are dependent on others, whereas some are independent. Employees of game development organizations or studios were selected as the targeted respondents of this study. In this study, the term "developer" is used to refer to any game development team member. For purposes of data collection, the authors initially joined various game development community forums.

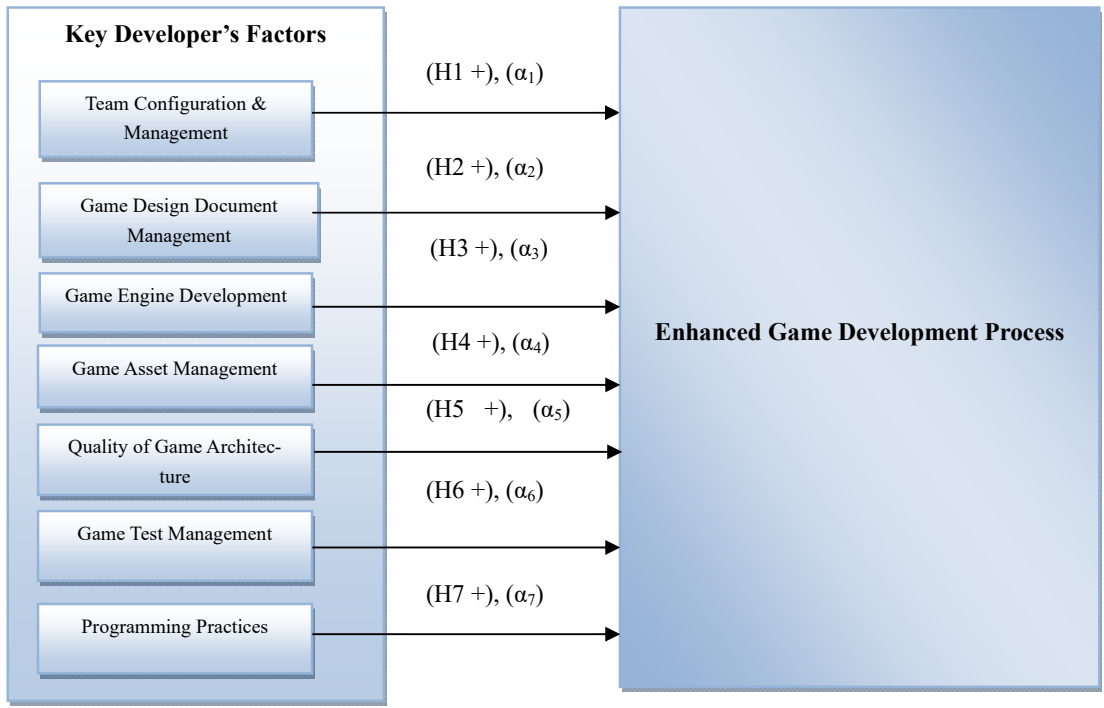

Fig 1. Research model

The respondents participating in the study were part of multinational organizations in Asia, Europe, and North America; statistics describing them are presented in Fig. 2

The organizational participants agreed to take part in the study based on a mutual agreement that their identities would be kept confidential. The size of the game project development teams varied from 10 to 50. Fig. 3 shows the total time period of the game development projects considered by respondents while answering the measuring instrument. Figure 4 represents the number of respondents based on their development role in the game project. Figure 5 shows the percentage of development methodologies used by respondents for any particular game project.

The participants in the study were mainly part of game projects that were developed for different platforms such as kiosks and standalone devices, the Web, social networks, consoles, PC/Macs, and mobile phones. The game genres implemented in most of their projects included action or adventure, racing, puzzles, strategy/role playing, sports, music-based, and other categories. The qualifications for this study were that the respondent must be a part of a development team that had at least three full-time developers; that the respondent worked on the project for at least one- third of its total duration;

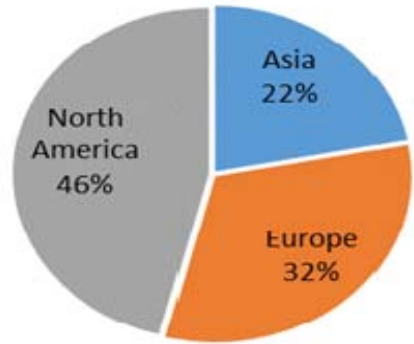

Fig. 2 Number of respondents by continent

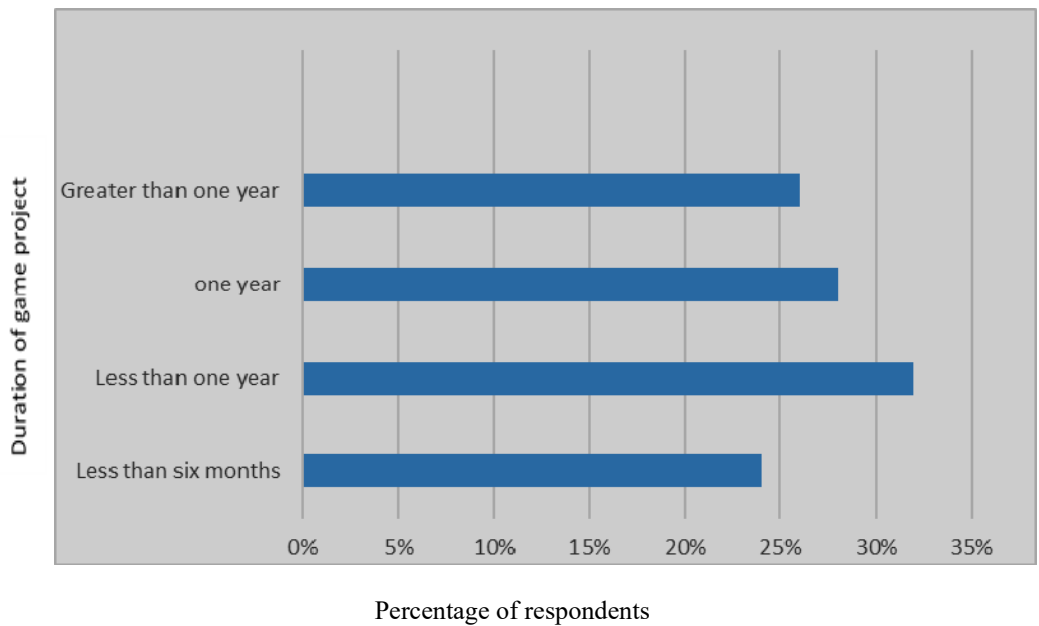

Fig. 3 Total software game development duration of particular game projects considered by respondents.

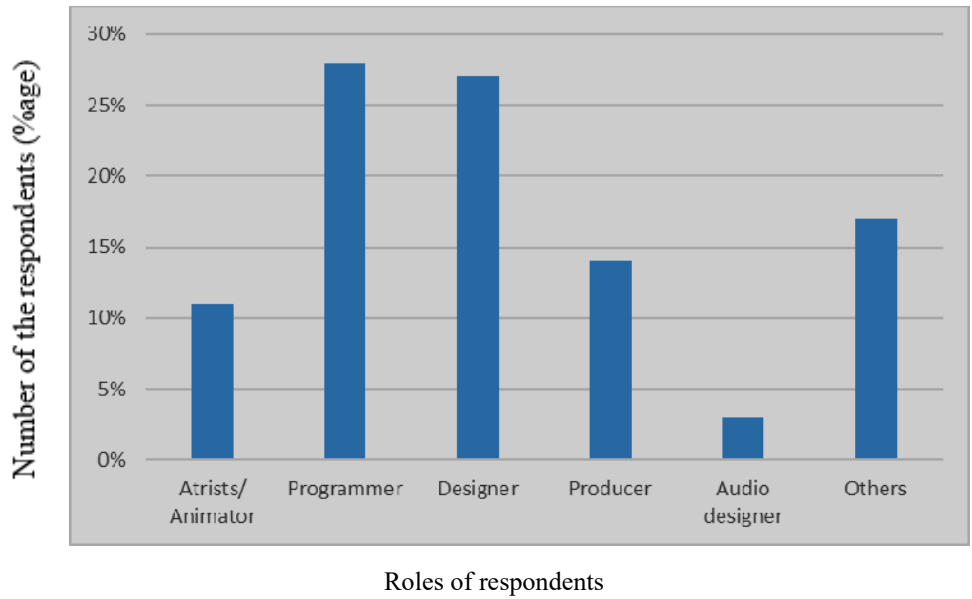

Roles of respondents

Fig. 4 Percentage of respondents based on their role in the development process.

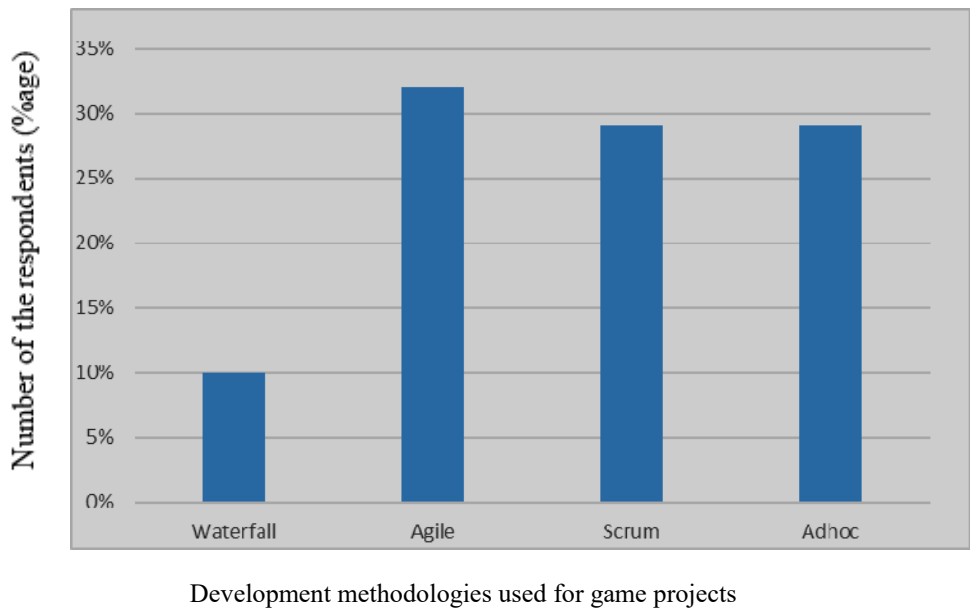

Development methodologies used for game projects

Fig. 5 Percentage of development methodologies used by respondents.

and that the project was either completed or cancelled within the last three years.

Finally, respondents must have worked in the development team in some sort of development role, such as a designer, artist, animator, programmer, producer, or sound designer. The survey respondents worked in various capacities such as game designer, artist, programmer, audio designer, and producer. The total number of survey respondents was 118, including a minimum of one and a maximum of four responses from each organization. Although the collected sample size is relatively large but it is still considered small sample as compared to the population size.

**4.1 Measuring instrument**

This study gathered data on the key developer's factors and the perceived level of enhanced game development process identified in the research model depicted in Fig. 1. To learn about these two topics, the questionnaire presented in Appendix A was used as a data collection instrument. First, organizations involved in the game development process were asked to what extent they practiced the identified key developer factors for the game development project in question. Second, they were asked what they thought of the enhanced game development process for different games in the software game industry. The five-point Likert scale was used in the questionnaire, and with each statement, the respondents were required to specify their level of agreement or disagreement. Thirty-four items were used to measure the independent variables (the key factors), and for the dependent variable (enhanced game development process), nine items were used. The literature related to key developer's factors was reviewed in detail to ensure a comprehensive list of measurement items for each factor from the literature. A multi-item, five-point Likert scale was used to measure the extent to which each key developer factor was practiced for the game development project. The Likert scale ranged from (1) meaning "strongly disagree" to (5) meaning "strongly agree" and was associated with each item. The items for each identified factor were numbered from 1 to 34 in Appendix A and also labelled sequentially. They were measured for each project that was completed within the last three years based on a multi-item five-point Likert scale. The enhanced game development process was the dependent variable, and designated items for the dependent variable were numbered separately from one to nine and labelled sequentially. All the items specifically written for this study are presented in Appendix A. To the best of the authors' knowledge, this is the first empirical study of key

Table 2 Cronbach's alpha coefficient and principal component analysis of seven variables.

| Developers factor | Item no. | Coefficient α | PC eigenvalue |
|---|---|---|---|
| *Team Configuration & Management* | 1–6 (excluded 1) | 0.63 | 1.48 |
| *Game Design Document Management* | 7–11(excluded 10) | 0.60 | 1.51 |
| *Game Engine Development* | 12–18(excluded 18) | 0.68 | 1.49 |
| *Game Asset Management* | 19–22(excluded 22) | 0.81 | 1.57 |
| *Quality of Game Architecture* | 23–25 | 0.84 | 1.01 |
| *Game Test Management* | 26–29 | 0.64 | 1.79 |
| *Programming Practices* | 30–34(excluded 30) | 0.86 | 1.25 |

software game developer's factors for the enhanced game development process in the SGI.

### 4.2 Reliability and validity analysis

To perform reliable and valid research, quantitative analysis was carried out. Two integral measure of precision, reliability and validity analysis, were used to conduct empirical studies. The consistency or reproducibility of a measurement is referred to as reliability. On the other hand, valid inference or agreement between the measured and true values is referred to as validity. The measuring instrument designed for this study was also tested by reliability and validity analysis. The test was based on common practices usually used for empirical analysis. Reliability analysis was performed to determine the internal consistency of the multi-scale measurement items designed for the seven identified factors. To evaluate internal consistency, Cronbach's alpha [95] coefficient was used. Criteria for Cronbach's alpha ranging from 0.55 to 0.70 were considered satisfactory. Researchers have reported different ranges of satisf-

-actory criteria for Cronbach's alpha based on their findings. Osterhof [96] suggested that a value of 0.60 or higher was satisfactory for reliability coefficients based on his findings. Nunnally and Brenste [97] reported that a value of 0.70 or higher for a reliability coefficient can be considered satisfactory for any measuring instrument.

Van de Ven and Ferry [98] recommended that a value of 0.55 or higher of the reliability coefficient could be considered satisfactory. A first calculation was performed on a sample dataset to determine the reliability of the dataset using

Cronbach's alpha coefficient. Some of the assessment items for each factor were excluded if they affected the desired value of Cronbach's alpha coefficient. In the sample dataset, other than item no. 1 of team & configuration management, item no. 10 of game design document management, item no. 18 of game engine development, item no. 22 of game asset management and item no. 30 of programming practices, all assessment items were found reliable. So, we removed item no. 1, item no. 10, item no. 18, item no. 22 and item no. 30 from the instrument. After this, the whole dataset was evaluated using Cronbach's alpha coefficient. The results of these calculations showed that reliability coefficients for the seven factors ranged from 0.61 to 0.76. These coefficients are reported in Table 2. Hence, all variables developed for this study could be considered reliable.

Validity analysis was performed for the dataset using principal component analysis (PCA) [99]. PCA is usually used for convergent validity analysis and was calculated here for seven factors. Campbell and Fiske [100] suggested that convergent validity has occurred in a given case only if the scale items in a measurement instrument are highly correlated and if they move in the same direction in a given assembly. The construct validity of PCA-based analysis was determined using the eigenvalue criterion [101]. Here, a criterion value greater than one was used to retain any component based on the Kaiser criterion [102]. Eigenvalue analysis showed that out of the seven variables, five together formed a single factor, whereas game design document management and programming practices loaded on a second factor, and both eigenvalues were greater than one. The reported convergent validity of this study was considered adequate.

### 4.3 Data analysis techniques

To perform the empirical investigation for this study, various statistical approaches were used. Initially, the research activity was divided into three phases to evaluate the significance of the proposed hypotheses H1–H7. In phase I, parametric statistical and normal distribution tests were performed. A non-parametric statistical approach was used in phase II, and for the analysis, a Partial Least Squares (PLS) analysis was carried out.

To address external threats to validity, both parametric and non-parametric approaches were used. Parametric approach is used to measures the strength of the linear relationship between normally distributed variables. When the relationship between the variables is not linear or the variables are not normally distributed then it may be more appropriate to use the non-parametric approach. Due to the small sample size, both parametric and nonparametric approaches were used to address the threat to external validity and we found results of both approaches are consistent.

The measuring instrument contains multiple items for each independent and dependent variable, and respondent ratings were aggregated to obtain a composite value. Using a parametric statistical approach in phase I, the Pearson correlation coefficient was calculated for the tests, with a one-tailed *t*-test for each hypothesis H1–H7. For phase II, the Spearman correlation coefficient was used to test hypotheses H1–H7 using a non-parametric statistical approach. Phase III of the empirical investigation was carried out to address issues of non-normal distribution and complexity or small sample size of the dataset. Fornell and Bookstein [103] and Joreskog and Wold [104] reported that if non-normal distribution, complexity, small sample size, and low theoretical information are issues, then partial least squares (PLS) analysis will be helpful.

The PLS technique was used in Phase III to increase the reliability of the results and deal with the limitation of small sample size. For statistical calculations, the Minitab 17 software was used.

## 5. Data Analysis and Results

### 5.1 Phase I of hypothesis testing

To test hypotheses H1–H7, parametric statistics were used in this phase. The Pearson correlation coefficient was determined between the independent variables (developer's factors) and the dependent variable (the enhanced game development process) of the research model, as illustrated in Fig. 1. The level of significance to accept or reject the hypotheses was then selected. Each hypothesis was accepted if its *p*-value was less than 0.05 and rejected if its *p*-value [105] was greater than 0.05. In Table 3, calculated results for the Pearson correlation coefficient are listed.

Table 3 Hypothesis testing using parametric and non-parametric correlation coefficients.

| Hypothesis | Key factors | Pearson correlation coefficient | Spearman correlation Coefficient |
|---|---|---|---|
| H1 | Team configuration and management | 0.29* | 0.29* |
| H2 | Game design document management | 0.79* | 0.74* |
| H3 | Game engine development | 0.59* | 0.64* |
| H4 | Game asset management | 0.45* | 0.47* |
| H5 | Quality of game architecture | 0.13** | 0.19** |
| H6 | Game test management | 0.42* | 0.37* |
| H7 | Programming practices | 0.52* | 0.48* |

*Significant at *p*<0.05                                                                                   **Insignificant at *p*>0.05

Hypothesis H1 was accepted because the Pearson correlation coefficient between team configuration and management and the enhanced game development process was positive (0.29) at $p<0.05$. For hypothesis H2 concerning game design document management and the enhanced game development process, the Pearson correlation coefficient was also positive (0.79) at $p<0.05$, and therefore hypothesis H2 was also accepted. Hypothesis H3 concerning game engine development and the enhanced game development process was accepted due to a positive (0.59) correlation coefficient at $p<0.05$. Hypothesis H4 concerning game asset management and the enhanced game development process was accepted based on its positive Pearson correlation coefficient (0.45) at $p<0.05$. Hypothesis H5 concerning quality of game architecture and the enhanced game development process was rejected based on its positive correlation coefficient (0.13), but higher $p>0.05$.

Hypothesis H6 regarding game test management and the enhanced game development process was accepted due to its positive Pearson correlation coefficient (0.42) at $p<0.05$. The last hypothesis (H7) relating programming practices to the enhanced game development process was also found to be significant (0.52) at $p<0.05$ and was therefore accepted. Hence, in summary, hypotheses H1, H2, H3, H4, H6, and H7 were accepted and found to be statistically significant. Hypothesis H5 was not supported statistically and was therefore rejected.

### 5.2 Phase II of hypothesis testing

Hypotheses H1–H7 were tested based on the non-parametric Spearman correlation coefficient in phase II. Table 3 reports the results for the Spearman correlation coefficient. Hypothesis H1 regarding team configuration and management was accepted because of its positive Spearman correlation coefficient (0.29) at $p<0.05$.

Table 4 PLS regression results for hypothesis testing

| Hypothesis | Factors | Path coefficient | $R^2$ | $F$-Ratio |
|---|---|---|---|---|
| H1 | Team configuration and management | 0.29 | 0.08 | 11.35* |
| H2 | Game design document management | 0.74 | 0.56 | 148.9* |
| H3 | Game engine development | 0.59 | 0.34 | 62.09* |
| H4 | Game asset management | 0.07 | 0.006 | 0.72* |
| H5 | Quality of game architecture | 0.13 | 0.02 | 2.3** |
| H6 | Game test management | 0.42 | 0.18 | 26..20* |
| H7 | Programming practices | 0.52 | 0.27 | 44.45* |

*Significant at $p<0.05$      **Insignificant at $p>0.05$

The Spearman correlation coefficient for game design document management and the enhanced game development process (hypothesis H2) was also positive (0.74) at $p<0.05$ and was also found to be significant. The relationship between game engine development and the enhanced game development process game (hypothesis H3) was found to be statistically significant due to its Spearman correlation coefficient (0.64) at $p<0.05$ and was accepted. For hypothesis H4 regarding game asset management, the Spearman correlation coefficient was positive at $p<0.05$, and therefore H4 was accepted. Hypothesis H5 concerning quality of game architecture and the enhanced game development process was rejected due to its positive coefficient (0.19) at $p>0.05$.

Hypothesis H6 concerning game test management and the enhanced game development process was accepted due to its positive Spearman correlation coefficient (0.37) at $p<0.05$. The last hypothesis (H7) relating programming practices to the enhanced game development process was also found to be significant (0.48) at $p<0.05$. In summary, hypotheses H1, H2, H3, H4, H6, and H7 were accepted and found to be statistically significant. Hypothesis H5 was not supported statistically and was therefore rejected.

### 5.3 Phase III of hypothesis testing

Hypothesis testing in phase III was performed using the partial least squares (PLS) technique. The main reason for using the PLS method in this phase was to cross-validate the results obtained from the parametric and non-parametric statistical approaches used in Phases I and II and to overcome their associated limitations.

Tests were also per formed on hypotheses H1–H7 to check their direction and significance. The dependent variable, i.e., the enhanced game development process, was designated as the response variable and other individual factors (independent variables) as the predicate variables for PLS examination. The observed results of the structural hypothesis tests are presented in Table 4. The table also includes the values of the path coefficient, $R^2$, and the $F$-ratio. The path coefficient for team configuration and management (H1) was observed to be 0.29, with an $R^2$ of 0.08 and an $F$-ratio of 11.35, and H1 was therefore found to be significant at $p<0.05$. Game design document management (H2) had a positive path coefficient of 0.74, $R^2 = 0.56$, and $F$-ratio = 148.9 and was also found to be statistically significant at $p<0.05$. Game engine development (H3) had a path coefficient of 0.59, a very low $R^2$ of 0.34, and an $F$-ratio of 62.09 and was found to be significant at $p<0.05$. Game asset management (H4) had a positive path coefficient of 0.07, a low $R^2$ of 0.06, and an $F$-ratio of 0.72 and was judged to be significant because the $p$-value was less than 0.05. Quality of game archit--ecture (H5) (path coefficient: 0.13, $R^2$: 0.02, and $F$-ratio: 2.3) was found to be statistically insignificant at $p<0.05$. Game test management (H6) (path coefficient: 0.42, $R^2$: 0.18, and $F$-ratio: 26.20) and programming practices (path coefficient: 0.52, $R^2$: 0.27, and $F$-ratio: 44.45) were found to be significant at $p<0.05$.

### 5.4 Research model testing

The linear regression equation for the research model is given by Eq. 1. The research model was tested to provide empirical evidence that factors important to game developers play a considerable role in improving the overall game development process in the SGI. The test procedure examined the regression analysis, the model coefficient values, and the direction of the associations. The dependent variable (the enhanced game development process) was designated as the response variable and the other independent variables (all the key developer factors) as predicate variables. The regression analysis model results are reported in Table 5. The path coefficients of six of the seven variables (team configuration and management, game design document management, development of a game engine, game asset management, game test management, and programming practices) were positive and were found to be statistically significant at $p<0.05$. The path coefficient for quality of game architecture was positive, but was found not to be statistically significant at $p<0.05$. The overall $R^2$ value of the research model was 0.83, and the adjusted $R^2$ value was 0.68 with an $F$-ratio of 36.97, which was significant at $p<0.05$.

Table 5 Linear regression analysis of the research model

| Model coefficient name | Model coefficient | Coefficient value | t-value |
|---|---|---|---|
| **Team configuration and management** | $\alpha_1$ | 0.06 | 1.14* |
| **Game design document management** | $\alpha_2$ | 0.50 | 7.44* |
| **Game engine development** | $\alpha_3$ | 0.31 | 5.19* |
| **Game asset management** | $\alpha_4$ | 0.21 | 0.38* |
| **Quality of game architecture** | $\alpha_5$ | 0.03 | 6.57** |
| **Game test management** | $\alpha_6$ | 0.13 | 2.24* |
| **Programming practices** | $\alpha_7$ | 0.10 | 1.58* |
| **Constant** | $\alpha_0$ | 0.01 | 1.13* |
| **$R^2$** | 0.83 | Adjusted $R^2$ | 0.68 |
| **$F$-ratio** | 36.97* | | |

*Significant at $p<0.05$   **Insignificant at $p>0.05$

## 6. Discussion

Software game development is a multidisciplinary activity that has its roots in the management and software engineering disciplines. The software game industry has become a mass phenomenon, supplemented by a number of possible strategies and exciting questions for game development companies. More and more companies are entering the market, and hence the intensity of competition is increasing. Established and new entrants must both pay attention to the key factors that help to improve their game development processes and keep them competitive in the market. Now it is time to understand the perspective of game developers and to learn what they think is important to improve software game quality and how the developed game can become successful in the market. This research is a first step towards this understanding because it will help developers and game development organizations to understand the relationships and interdependences between key factors from a developer's perspective and to understand the enhanced game development process. This research is the first empirical investigation of factors important to developers in relation to improving the current development process and provides an opportunity to explore associations between them empirically. The observed results support the theoretical assertions made here and provide the very first evidence that consideration of key developers' factors while developing games is important for software game success. This could well result in institutionalizing the software game development approach, which in turn has a high potential to maximize profits.

Especially in the game development process, multidisciplinary team configuration and management is a huge challenge. Basically, producing high-quality games relies on a high level of planning, communication, and organization of multidisciplinary teams to avoid costly delays and failures. Many factors have been identified by researchers as important to implementing a successful collaboration between any kinds of multidisciplinary team. These factors include interpersonal factors such as trust among team members and ability to communicate [34], willingness to collaborate, and mutual respect [106]. Others are organizational factors, including establishing appropriate protocols and supporting collaboration [107] These factors can be implemented by using various software applications that are specifically designed for collaborating on commercial software development projects. The main concern when using these software applications is that they must fit in with the existing computing and workflow environment [108]. Management of the members of various multidisciplinary teams can be evaluated and maintained mainly by examining values and practices, for example, what each individual team member brings to the table, how they use material or assets produced by other team members, how they reconcile conflicting priorities, and finally how their personal relations influence the collaboration. The multidisciplinary team can use management or collaboration software for task tracking, version control, file sharing, and continuous integration. Successful collaboration between team members enables them to manage easily all phases of game development from start-up, creating a concept, creating a proof of concept, the production phase, and so forth until the game is published.

This study has explored the importance of team configuration and management factors from a developer's perspective. It has found positive associations between team configuration and management and the enhanced game development process. Hence, proper configuration and management of multidisciplinary teams is a crucial part of the game development process. However, it must be balanced with other development issues in the game development process.

Game design document management has been found to be positively associated with the enhanced game development process. The GDD is mainly a pre-production artifact which is defined by the pre-production phase team to capture a creative vision of the game. Game developers generally feel that imposing too much structure at the start of a game may be highly detrimental [40], resulting in reduced creativity, constraining expression, and risking the intangibles that create an enjoyable feeling or experience. At the same time, the importance of structure has been highlighted by many researchers, as discussed in the literature review section. Management of the game design document and its transition into a requirements and specifications document is challenging.

One way to handle this during the pre-production phase is to produce two documents. The first one is the GDD, and the second one is a requirements and specifications document based on the GDD. Managing and transforming the GDD into a production document is complex because the two require different documentation styles. Supportive documentation is also required to help the development team in its transition from pre-production to the production phase. The author of the GDD may not have the requisite writing skills to produce a document that is understandable by the production team (technical people). Basically, there is a long list of required skills for a GDD developer, such as knowledge of game design, technical communication, and requirements engineering. Hence, a formal process is needed to support the transition and would likely increase the reliability of the game development process. The results of this study have shown that development and transformation of the GDD is very important and also requires strong management skills to reduce documentation effort. Hence, the results presented here have shown that a good GDD is the greatest contributor to the success or failure of a game development project.

Game asset management was also found to have a positive association with the enhanced game development process. Game assets, defined as any piece of data that is in a format that can be used by the game engine, will be presented to the user. To create and manage game assets, a realistic content generator must be developed that can fill in the missing bits. Trade-offs between realism and performance and between realism and control must also be investigated for any asset created. For graphical animation, a number of 3D model formats can be used by game developers. These can generally be divided into two categories: frame-based animation and skeletal-based animation. Determination of the perfect animation model for a game has become crucial because a diversity of format types for graphics are available. Eventually, a poor choice could limit the performance of the game itself. For sound effects in games, certain problems are faced by developers because of unexpected or complex scene configurations. A number of asset management tools exist, but selecting the appropriate one is a challenge because each has its own limitations and benefits.

Improvements in the game development process have been greatly aided by the emergence of game development tools, specifically game engines. A game engine facilitates the game development process by providing various sets of features that help decrease development time and cost. These are available for most game genres (e.g., role-playing games or serious games for training) and vary in cost and complexity. Not all game engines support the entire feature set of all the game genres. Hence, integrating all the technological aspects into one framework is a prohibitively difficult task. It is understandable, therefore, that confusion exists among game developers with regard to selecting the appropriate game engine. Game development tools should be selected only after determining the game concept and the GDD [109]. Most researchers in the area of game development tools have proposed their own architectures for specific genres and platforms. Anderson *et al.* [110] raised some important open questions for the academic community that are specific to the game engine development research field. The first is the main issue of the lack of a development language. The second question is how to define the boundaries between the game loop and the game engine. For example, what technical aspects should a game engine cover in a game? The third problem is that there is no standardization for game engines because most of them are specific to a particular game genre and game project. The fourth issue involves design dependencies, and the last the need for best practices when creating game engines. It was generally agreed that a game engine should handle a diversity of inputs and outputs, a restricted set of customizations based on each genre, and an asset and resource management system. The results of this study have also showed that development of a game engine has a positive impact on the enhanced game development process. In other words, game engine development is an important factor that needs more consideration from a developer perspective.

It is a common perception that a good-quality or even perfect game architecture is a very important part of the game development process because reworking architecture afterwards is always hard. A game architecture identifies the main structural components of the underlying software and their relationships. In the game development literature, many researchers have proposed different frameworks for different platforms and based on different technologies. As a developer it is difficult to select among these because all provide a kind of reference architecture and their validity is still in question. The findings of the study do not support a statistically positive relation between quality of game architecture and the enhanced game development process. The direction of association was found to be positive, but the required level of confidence was not supported. Hence, the hypothesis that quality of game architecture has a positive impact on the enhanced game development process was statistically rejected.

Testing in game development is done mainly at a very late stage or the end of development to ensure the quality and functionality of the finished product. Typically, in a particular game project, the leader dedicates a specific amount of time for quality assurance or a beta tester to test the game. Various development methodologies are used to develop games, such as the agile methodology and the waterfall model, but testing must form part of all processes. Every aspect of a game should be tested during the development and production phases. In addition, certain foundational elements should also be tested during the pre-production phase, such as frameworks and platform set-up. The most important aspect of testing for game developers is to integrate testing as part of the production phase to improve efficiency. To ensure continuous quality and delivery of good games to the market, developers must consider majority testing options during the production phase. Helppi [80] also researched the possibility that mobile game robustness can be improved by continuous integration, delivery, and testing and concluded that this approach can improve the outcome of games and result as a more robust end-product. Therefore, testing plays an important role in each step of the development phase, and its management throughout the game development process is important. The results of this study have also supported the hypothesis that game testing management is important for the enhanced game development process. At the same time, testing techniques have matured over time, but still need improvement.

Game programming strategy has a direct effect on game performance. There are many concerns associated with today's game programming practices. Game developers must look for solutions to common problems in game programming such as coupling of modules, availability of different scripting and programming languages, platform compatibility issues, memory management, and code optimization strategy, specifically to improve game performance and quality. Hence, game developers must consider various aspects of the game such as speed, flexibility, portability, and maintainability while still coding. Ultimately, the skilled programming team will be able to develop and implement the full functional game. Matching of required skills to the abilities of developers is very important to improve the overall game development process, a conclusion also supported by this study.

Overall, the findings of this study are important for the development of good-quality software games. Rapid and continual changes in technology and intense competition not only affect the business, but also have a great impact on development activities. To deal with this strong competition and high pressure, game development organizations must continually assess their activities and adopt an appropriate evaluation methodology. Use of a proper assessment methodology will help the organization identify its strengths and weaknesses and provide guidance for improvement. However, the fragmented nature of the game development process requires a comprehensive evaluation strategy, which has not yet been entirely explored. The findings of this study will help game development organizations to look for contributing key success factors from a developer's perspective. This study is a part of a larger project aiming to propose a software game maturity assessment model. The developer's perspective was one of the important dimensions identified among the consumer, the business [111], and the process itself. The findings of this study also provide a justification to include these factors in the process assessment methodology.

### 6.1 Limitations of the study

For software engineering processes or product investigations, various empirical approaches are used, such as case studies, metrics, surveys, and experiments. However, certain limitations are associated with empirical studies and with this study as well. Easterbrooks *et al.* [112] suggested four criteria for validating empirical studies: internal validity, construct validity, external validity, and reliability. Wohlin *et al.* [113] stated that generalizing experimental results to industrial practice by researchers is mostly limited by threats to external validity. In this study, measures were taken to address external threats to validity. The random sampling method was used to select respondents from all around the world. Open-ended questions were also included in the questionnaire.

The choice and selection of independent variables was one of the limitations of this study. To analyze the association and impact of factors affecting software game success, seven independent variables were included. However, other key factors may exist which have a positive association with and impact on the game development process, but due to the presence of the selected seven variables in the literature, they were included in the study. In addition, other key factors may exist, such as regionally or environmentally based choices, which may have a positive impact on the game development process, but were not considered in this study. Furthermore, the focus of this study was only on developers' factors affecting the enhanced game development process.

Another notable limitation of the study is the small sample size. Although, collected number of responses are large in number but they can still consider small as compared to overall population size. The vast majority of game developers work in one- to three-person teams and did not have the required level of experience (three years) and were therefore excluded from this empirical investigation. Most respondents refused to answer the questionnaire because they were too busy in the game development process or launching their games in the market. Therefore, data collection from the game industry was limited, resulting in small sample size. There are some approaches discussed by researchers such as by Zhang and Zhang [117] to handle the small sample size by using different machine learning techniques. However, one of the objective to divide data analysis section into three phases is to address the small sample size issue. The main effect of small sample size is on its statistical power, Type II error, significance and on distribution [115]. Therefore, the important thing is while making conclusion avoid strong statements. As, the small samples size studies results can be difficult to replicate or generalize [116] but they do provide some interplay between variables. The well designed small studies are seems ok to conduct as they provide quick results but they need to be interpreted carefully [114]. The results of small studies should be used to design larger confirmatory studies which is the case of this study as well.

In software engineering, the increased popularity of empirical methodologies has raised concerns about ethics. This study has adhered to all applicable ethical principles to ensure that it would not violate any experimental ethics guidelines. Regardless of its limitations, this study has contributed to the software game development process and has helped game development organizations understand the developer's dimension of software games.

## 7. Conclusions

Game development is a complex process, and for successful development of good-quality software games, game developers must consider and explore all related dimensions as well as discussing them with all the stakeholders involved. This study provides a better understanding of the factors important to developers in the software game development process and explores the impact of key factors on the success of software games from a developer's perspective. This study has mainly tried to answer the research question that was posed earlier in this paper and to analyze the impact of developers' key factors for game develop-ment process improvement. The results of this empirical investigation have demonstrated that developers' key factors are very important and play a key role in improving the software game development process. The results showed that team configuration and management, game design document management, game engine development, game test management, and programming practices are positively associated with the enhanced game development process. The empirical investigation found no strong association or impact between quality of game architecture and the enhanced game development process. In the game development field, this research is the first of its kind and will help game developers and game development organizations achieve a better understanding of key factors for improving the game development process. To improve the current game development process and develop good-quality games, it is important for developers to consider the identified key factors as well as others. Currently, the authors are working on developing a software game maturity model for game development process assessment. This study has provided the empirical evidence and justification to include factors from the developers' perspective in evaluating the developer dimension of game development process maturity.

**Dr. Saiqa Aleem** received her MS in Computer Science (2004) from University of Central Punjab, Pakistan, MS in Information Technology (2013) from UAEU, United Arab Emirates and PhD. in Software Engineering (2016) from University of Western Ontario, Canada. She had many years of academic and industrial experience holding various technical positions. She is Microsoft, CompTIA, and CISCO certified professional with MCSE, MCDBA, A+ and CCNA certifications.

**Dr. Luiz Fernando Capretz** has vast experience in the software engineering field as practitioner, manager and educator. Before joining the University of Western Ontario (Canada), he worked at both technical and managerial levels, taught and did research on the engineering of software in Brazil, Argentina, England, Japan and the United Arab Emirates since 1981. He is currently a professor of Software Engineering and Assistant Dean (IT and e-Learning), and former Director of the Software Engineering Program at Western. His current research interests are software engineering, human aspects of software engineering, software analytics, and software engineering education. Dr. Capretz received his Ph.D. from the University of Newcastle upon Tyne (U.K.), M.Sc. from the National Institute for Space Research (INPE-Brazil), and B.Sc. from UNICAMP (Brazil). He is a senior member of IEEE, a distinguished member of the ACM, a MBTI Certified Practitioner, and a Certified Pro-fessional Engineer in Canada (P.Eng.). He can be contacted at lcapretz@uwo.ca; further information can be found at:

http://www.eng.uwo.ca/people/lcapretz/

**Dr. Faheem Ahmed** received his MS (2004) and Ph.D. (2006) in Software Engineering from the Western University, London, Canada. Currently he is Associate Professor and Chair at Thompson Rivers University, Canada. Ahmed had many years of industrial experience holding various technical positions in software development organizations. During his professional career he has been actively involved in the life cycle of software development process including requirements management, system analysis and design, software development, testing, delivery and maintenance. Ahmed has authored and coauthored many peer-reviewed research articles in leading journals and conference proceedings in the area of software engineering. He is a senior member of IEEE.

**Appendix A**

**Measuring Instrument**

This survey attempts to evaluate key success factors in the game development process statistically from a developer's perspective. This survey captures the opinions of game developers who have completed game projects regarding factor collaboration, game design documents, game engines, game asset creation, game architecture, game testing, and programming.

If you are a game developer with at least one team project under your belt, please help us by taking the survey below for the game project you completed most recently and also give your opinion about enhanced the game development process.

**Section I – Qualifying questions**

1. Please take this survey for the most recent game development project for which you can answer "yes" to ALL the following questions:

  ☐ There were at least three fulltime developers on this team.

  ☐ I worked on the project for at least one-third of its total duration.

  ☐ The project was either completed or cancelled sometime within the last three years.

  ☐ I worked in the development team in some sort of development role, such as a designer, artist, animator, programmer, producer, or sound designer.

This survey should not take more than 10–15 minutes to complete.

Your answers will be kept confidential, but the AGGREGATE data will be released to the public along with our conclusions.

**Section II: Background questions**

2. What is your region? \_\_\_\_\_\_\_\_\_\_\_\_\_\_\_

3. What was the total duration of your game development project? Enter the whole number.
Years: \_\_\_\_\_\_\_\_\_\_\_\_\_\_\_\_\_\_\_\_ Months: \_\_\_\_\_\_\_\_\_\_\_\_\_\_\_\_\_\_

4. Approximately, what is the size of development team?\_\_\_\_\_\_\_\_\_\_\_\_\_\_\_\_

**5. Please describe your primary role in the development process. Please select all that apply.**

Artist/animator ☐ Programmer ☐ Designer ☐ Producer ☐

Audio Designer ☐ Other (please specify) ☐

**6. The developed game was released for which platform? Select all that apply.**

Any desktop ☐ Any handheld device ☐ Any console ☐ Web ☐

Any mobiles ☐ Other ☐

**7. What was the genre of the developed game?**

Please Answer: ____________

**7. Which software development methodology was used to develop the game?**
Please pick the approach that seems closest based on the descriptions below.

☐ Don't know

☐ Waterfall: the project was divided into phases which included upfront planning, requirements, design, development, and testing phases.

☐ Agile: Project leaders evaluate the project priorities on weekly or monthly sprints. Iterative development was focussed on individual features, and frequent feedback was emphasized rather than requirements, specifications, or design documents.

☐ Agile using Scrum: the project followed the "Scrum" implementation of Agile. Priorities were determined by self-organizing cross-disciplinary teams. These teams were responsible for their own tasking and held daily scrum meetings to identify the work being done and bottlenecks to development.

☐ Other/Ad-hoc

**Section III:**

**Evaluation of enhanced game development process success factors identified through literature review**

The questionnaire objective is to find out which factors have a positive impact on the game development process. Please select the correct scale based on your best knowledge.

| key factors for the game development process from a developer's perspective | | | | | | | |
|---|---|---|---|---|---|---|---|
| Likert scale (1 = strongly disagree; 2= disagree; 3 = neutral; 4= agree; 5 strongly agree) | | 1 | 2 | 3 | 4 | 5 | N/A |
| **Team Configuration & Management** | | | | | | | |
| 1 | The team must be organized into sub-teams by discipline (art, programming, design) rather than by features. | | | | | | |
| 2 | Team members must have a similar vision of the game throughout the development process. | | | | | | |
| 3 | There must be support from lead management to the team members. | | | | | | |
| 4 | The entire team should be involved in prioritizing the work to be done for each milestone or sprint. | | | | | | |
| 5 | In case of any significant change in the game design or architecture, then all stakeholders must participate in the decision process. | | | | | | |
| 6 | The development plan for the game should be clear and well communicated to the team. | | | | | | |
| **Game Design Document Management** | | | | | | | |
| 7 | There must be a design document available to the team near the beginning of development that clearly specifies the game goals. | | | | | | |
| 8 | Priorities must be given to different components so the team will know which part is more important. | | | | | | |
| 9 | Details about storyboard, script writing, characters, and major and minor goals must be included in the GDD. | | | | | | |
| 10 | The GDD was understandable because it was well written. | | | | | | |
| 11 | Transformation of the GDD from the pre-production phase to the production phase was not problematic. | | | | | | |
| **Game Engine Development** | | | | | | | |
| 12 | The development platform and tools must be familiar to game developers. | | | | | | |
| 13 | The selected development tool provided asset and resource management. | | | | | | |
| 14 | The selected game engine was able to handle diverse type of input and output | | | | | | |
| 15 | Integration of all technological components was easy. | | | | | | |
| 16 | The game engine provided support for multi-platform development. | | | | | | |
| 17 | The development tool enables use of other embedded tools that are helpful in extension of current capabilities. | | | | | | |
| 18 | Reuse of the game engine is highly desirable. | | | | | | |
| **Game Asset Management** | | | | | | | |
| 19 | Realism and performance analysis must be a part of asset creation. | | | | | | |
| 20 | Realism and control investigation before asset creation is important. | | | | | | |
| 21 | Integration of sound effects into complex and unexpected scenes can usually be done by using available audio processing libraries. | | | | | | |
| 22 | Asset version control management must be performed to track different versions. | | | | | | |
| **Quality of Game Architecture** | | | | | | | |
| 23 | Gameplay was divided into different modules, and each module could be modified and tested independently without impacting other modules | | | | | | |
| 24 | Different game modules should be easily portable and extensible so that they can be plugged into other game projects. | | | | | | |
| 25 | The game architecture included robustness features that enable a game to be functional under unexpected circumstances. | | | | | | |
| **Game Test Management** | | | | | | | |
| 26 | Game testing steps were usually established during the pre-production phase and documented properly. | | | | | | |
| 27 | Game testing was performed throughout the game development process. | | | | | | |
| 28 | A suitable testing approach was selected to test game performance and quality. | | | | | | |

| | | | | | | |
|---|---|---|---|---|---|---|
| 29 | The game was tested for performance under various loads. | | | | | |
| **Programming Practices** | | | | | | |
| 30 | Programming team responsibilities and job roles were carefully matched with their particular programming skills and abilities. | | | | | |
| 31 | Programming style must be uniform among all programmers. | | | | | |
| 32 | Good commenting reduces the errors in code and speeds up the code review process. | | | | | |
| 33 | Standard naming and coding conventions should be used. | | | | | |
| 34 | Performance and optimization techniques (such as methodological and code optimization and datatype optimization) were applied to the code. | | | | | |
| **Enhanced game development process** | | | | | | |
| 1 | The game engine should allow rapid prototyping of new levels, behaviour, and scenarios and support dynamic content loading. | | | | | |
| 2 | Game architecture should be easy to understand, change, reuse, and debug. | | | | | |
| 3 | The game design document should be developed in a formal way and have all specifications such as executive summary, product, game and art specifications. | | | | | |
| 4 | Game assets should be created to fit into the game concept and must have a positive effect on the appearance of the game. | | | | | |
| 5 | Coding priorities must be established as a part of technical design and must be properly documented | | | | | |
| 6 | Before selection of a programming strategy, issues such as coupling between modules, performance, memory management, and availability of different programming paradigms should be taken into consideration. | | | | | |
| 7 | All aspects of the game were tested, such as game play, functionality, interaction control, connectivity issues, input controller, and platform compatibility. | | | | | |
| 8 | The entire team should meet frequently to openly discuss topics of interest, ask questions, and identify production bottlenecks. | | | | | |
| 9 | Game testing should be performed properly to ensure game performance and quality. | | | | | |